# Machine learning-based aerosol characterization using OCO-2 $O_2$ A-band observations

Sihe Chen[1,*], Vijay Natraj[2], Zhao-Cheng Zeng[1,3], and Yuk L. Yung[1,2]

[1]: Division of geological and planetary sciences, California Institute of Technology
[2]: Jet Propulsion Laboratory
[3]: Joint Institute for Regional Earth System Science and Engineering (JIFRESSE), University of California, Los Angeles

[*]: Corresponding author, email: sihechen@caltech.edu

Abstract
Aerosol scattering influences the retrieval of the column-averaged dry-air mole fraction of $CO_2$ ($XCO_2$) from the Orbiting Carbon Observatory-2 (OCO-2). This is especially true for surfaces with reflectance close to a critical value where there is very low sensitivity to aerosol loading. A spectral sorting approach was introduced to improve the characterization of aerosols over coastal regions. Here, we generalize this procedure to land surfaces and use a two-step neural network to retrieve aerosol parameters from OCO-2 measurements. We show that, by using a combination of radiance measurements in the continuum and inside the absorption band, both the aerosol optical depth and layer height, as well as their uncertainties, can be accurately predicted. Using the improved aerosol estimates as *a priori*, we demonstrate that the accuracy of the $XCO_2$ retrieval can be significantly improved compared to the OCO-2 Level-2 Standard product. Furthermore, using simulated observations, we obtain estimates of the error in the retrieved $XCO_2$. These simulations indicate that the bias-corrected OCO-2 Lite data, which is used for flux inversions, may have remaining biases due to misinterpretation of aerosol effects.

1.Introduction
The presence of aerosols has a profound influence on the global energy budget. These particulates constitute one of the major sources of uncertainty in the estimation of global climate sensitivities [1][2]. In addition, scattering effects due to aerosols make it hard to constrain the atmospheric pathlength. Indeed, imperfect characterization of atmospheric aerosols is the dominant error source in the retrieval of $CO_2$ from near-infrared space-based measurements [3]. The Orbiting Carbon Observatory-2 (OCO-2) measures the column-averaged dry-air mole fraction of $CO_2$ ($XCO_2$) from space using three near-infrared bands: the oxygen ($O_2$) A-band at 0.76 μm and the "weak" and "strong" $CO_2$ bands at 1.61 μm and 2.06 μm, respectively [4]. While methods such as direct observations of the solar disk reduce biases associated with aerosols, the presence of aerosols reduces the number of soundings available for retrieval if scattering effects are not properly accounted for [5]. In addition, it was shown that low-uncertainty aerosol priors reduces errors in near-infrared $CO_2$ retrievals [6]. The retrieved aerosol optical depth (AOD) from the OCO-2 full physics retrieval algorithm is very poorly correlated with AERONET AOD measurements [7]. In particular, for surfaces with reflectance near a critical value, the continuum radiance is not sensitive to changes in AOD, making aerosol characterization difficult. It was shown that near the critical surface albedo, XCO$_2$ retrievals suffer a significant loss of accuracy because the effects of changing the aerosol loading are identical to those due to changing the total absorbing gas column [8]. Figure 1 shows an illustration of the OCO-2 viewing geometry and a sample sounding in the O$_2$ A- and strong CO2 bands.

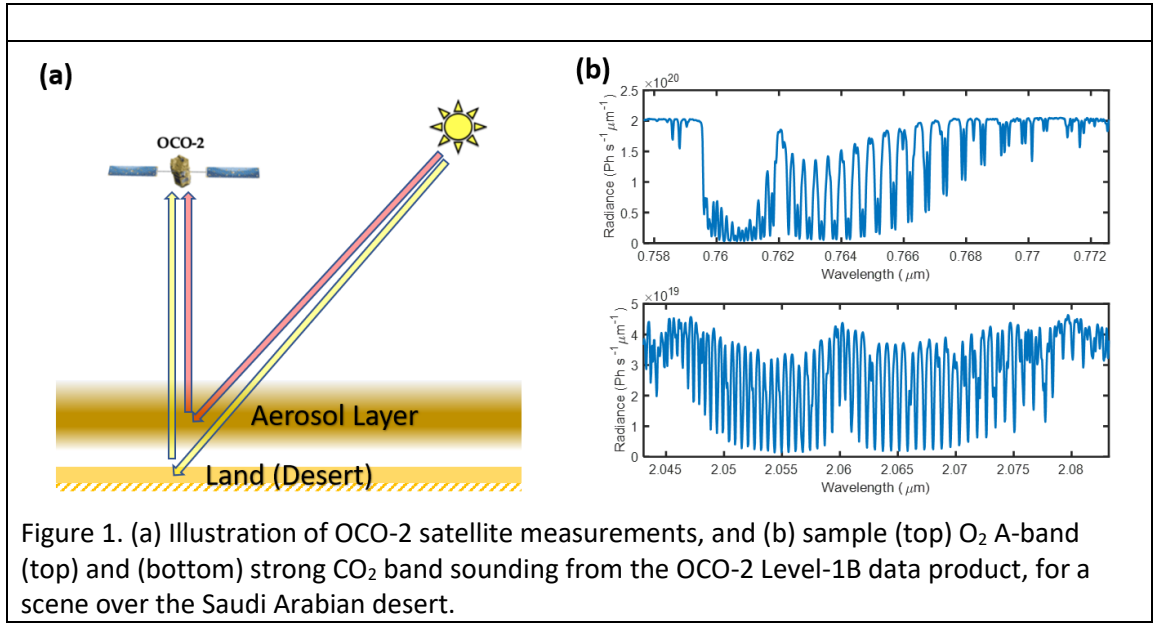

Figure 1. (a) Illustration of OCO-2 satellite measurements, and (b) sample (top) O$_2$ A-band (top) and (bottom) strong CO$_2$ band sounding from the OCO-2 Level-1B data product, for a scene over the Saudi Arabian desert.

To improve the characterization of aerosol vertical distribution, spectral sorting was appliedbased on the observed radiance, where different channels provide information about the AOD and the aerosol layer height (ALH) [9] [10]. There is a strong physical basis for the retrieval of aerosol information from the O$_2$ A-band. First, O$_2$ is homogeneously distributed in the atmosphere, and its abundance is well known. Second, the scattering effects of aerosols located at different altitudes have different signatures in different parts of the spectrum, allowing characterization of AOD and ALH [10].

Due to the variability of surface types, a similar characterization for soundings over land is expected to be more complicated. In particular, the critical albedo problem is illustrated in the left panel of Figure 2 for retrieving dust over deserts using O$_2$ A-band measurements. The critical surface albedo of ~0.4 in the continuum channels is equal to the reflectance of desert surfaces in this spectral region. However, this degeneracy can be resolved by using absorption channels (that seem to have a critical albedo as well, but at a different value).

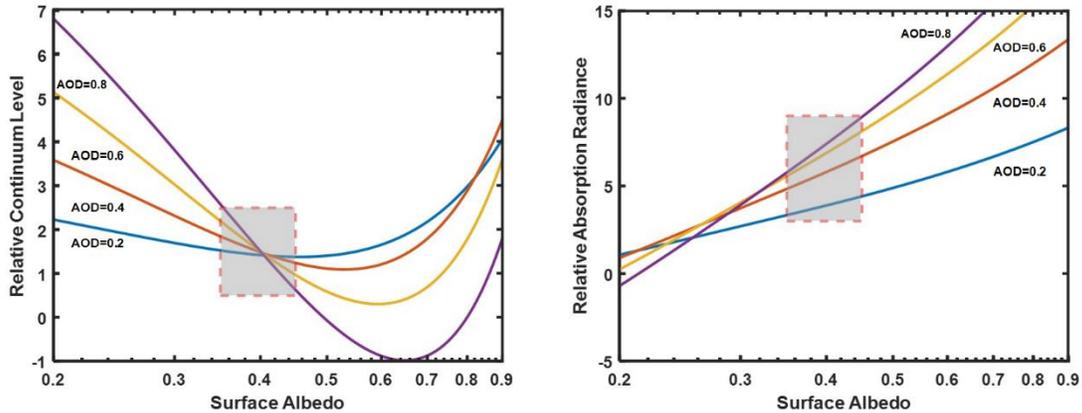

Figure 2. Sensitivity of $O_2$ A-band radiance in (left) continuum and (right) absorbing channels to AOD for different surface albedo values. Note that the x-axis is in log-scale. The results are obtained from simulations using the forward model in the version 8 ACOS algorithm [11].

While [10] correlated the radiance in absorbing channels with ALH via a linear relationship, a nonlinear method is required for more complicated scenarios such as over typical land surfaces. [12] used a neural network with 2557 inputs and a hidden layer with 50 to 500 neurons to produce highly accurate predictions of $XCO_2$ from simulated OCO-2 soundings, indicating the power of neural networks to capture the nonlinearity hidden inside OCO-2 measurements.

In this study, we use a machine learning approach to retrieve aerosol characteristics from OCO-2 observations. With a two-step classification-fitting neural network, we show that the AOD and ALH can be predicted accurately, providing a significant improvement compared to the OCO-2 Level-2 retrieval algorithm. Further, we show that by using a maximum likelihood loss function, the neural network delivers estimates of the uncertainties in the aerosol parameters. By using the improved estimates as *a priori* in the $XCO_2$ retrieval algorithm, we obtain $XCO_2$ with increased accuracy over surfaces with reflectance close to the critical albedo.

## 2. Data and Methods
### 2.1 OCO-2 spectra

We compress the full spectrum and extract only the most important information using the spectral sorting method [10]. Although this procedure incurs a loss of information, the improvement in the computational efficiency makes it much easier to handle the large OCO-2 data volumes while also providing more accurate AOD and ALH results.

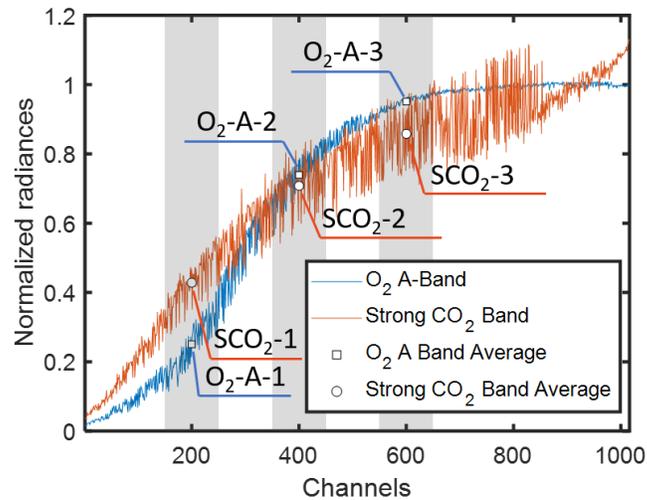

Figure 3. Illustration of spectral compression. The spectrum is sorted according to the absorption strength in clear-sky conditions. The grey bars denote the channels that are involved in the averaging, while the average values are marked and numbered. These average values constitute six of the nine inputs to the neural network.

We select three sets of 100 consecutive channels each in the $O_2$ A-band and the strong (2.06 µm) $CO_2$ band after the radiances are sorted; the average radiance of each set constitutes one input to the neural network. The choice of 100 consecutive wavelengths is based on a compromise between loss of information and compression of the input. An illustration of this procedure is shown in Figure 3. In addition to these six inputs, we include the continuum radiance in the $O_2$ A-band, the surface albedo, and the solar zenith angle as additional inputs to the neural network. Hence, for each sounding, the neural network has nine input variables. Since we only consider data from the nadir mode, the viewing zenith angle is not included as one of the variables. The AOD and ALH truth is calculated from Cloud-Aerosol Lidar and Infrared Pathfinder Satellite Observation (CALIPSO) Level-2 data [13], and the surface albedo is taken from the MCD43D3 dataset [14]. The data are compressed so that they do not contain spatial or temporal information, eliminating the risk of overfitting to nearby training data. We use the Python deep learning library PyTorch [15] to construct the neural networks in this work.

We expect the fitting to perform differently for different surface characteristics and aerosol types. This would entail the use of a large, high-quality training data set. If the dimensionality of the input is high, a huge neural network with many weights is required, which increases the risk of overfitting. By reducing the input dimension, a much smaller neural network is required, effectively preventing overfitting.

**2.2 Study area**

In this work, we focus on a region around the Riyadh megacity in Saudi Arabia. The observation area and a sample track is shown in Figure 4. We utilize data from five different days in 2016 for the OCO-2 satellite flyby near Riyadh. After removal of cloudy soundings, we keep 5000 different data points. We randomly select 45% of the data to form the training dataset, 10% to constitute the validation dataset and the remaining 45% to be used as the testing dataset.

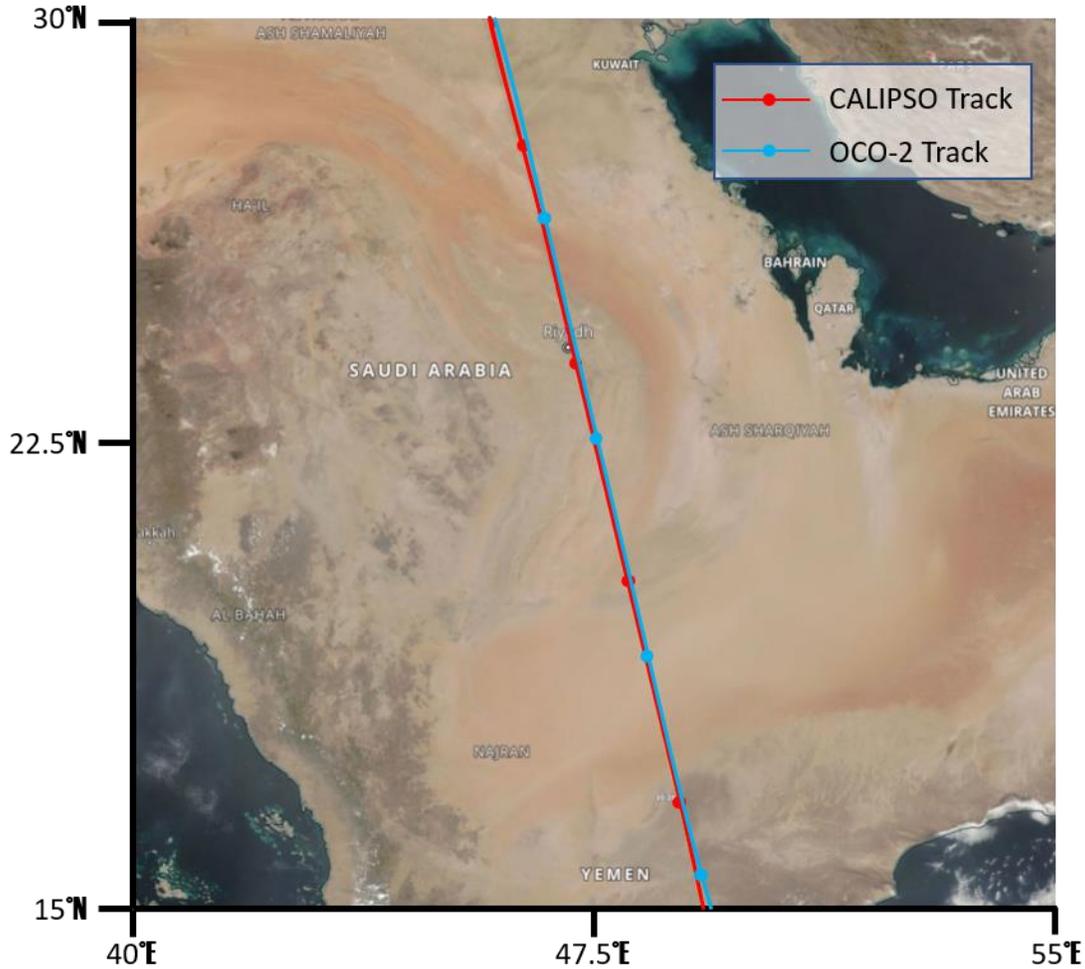

Figure 4. OCO-2 and CALIPSO tracks on May 22, 2016 over the Saudi Arabian Desert. The selected soundings have similar geometries, with the OCO-2 and CALIPSO ground tracks close to each other. The dots indicate the start and the end for data sections.

## 3. Machine learning
### 3.1 Training

The aim is to find an optimal neural network type and structure based on appropriate performance metrics. Two candidates are considered here: the first is the classical fully connected neural network (FC-NN), and the second is the specifically designed classification-fitting neural network (CF-NN).

The structure of FC-NN is shown in Figure 5. By trial and error, we find that 10 neurons in the hidden layer is sufficient and gives near-optimal performance under this structure.

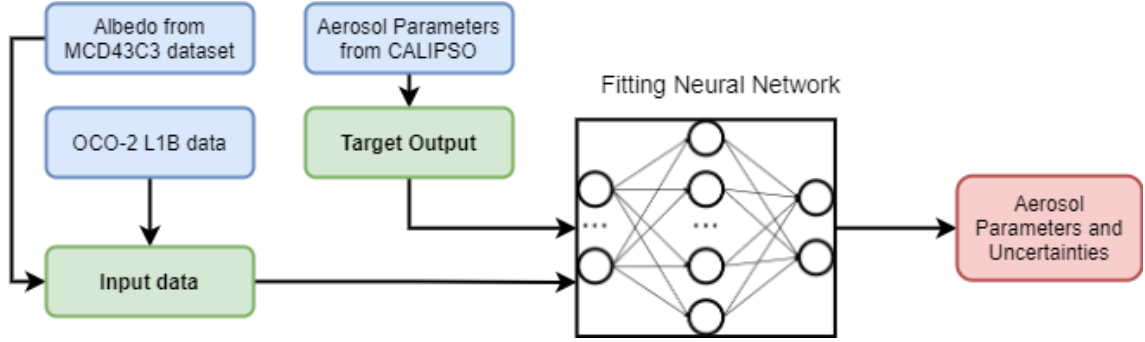

Figure 5. Structure of fully-connected neural network (FC-NN). The neural network contains a hidden layer of 10 neurons and an output layer containing AOD or ALH with uncertainty estimations.

On the other hand, CF-NN contains two sets of neural networks:

1. A classification network that finds which interval the AOD value falls into
2. A fully-connected fitting network that predicts the AOD and ALH

CF-NN is based on the hypothesis that the spectral behavior is different for small and large AOD. Hence, training separate models for these two scenarios is expected to improve prediction accuracy.

Since the training suffers from randomness and local minima, we use ensemble learning for both the classification and fitting steps. Out of 10 trained networks, we pick five for the classification layer with the best performance on the validation set, and choose the ensemble majority for the prediction. On the other hand, we pick the three best-performing networks for the fitting layer and take the weighted average as the prediction.

The AOD threshold values for the classification network are obtained by maximizing the information from each prediction. The average information *I* from each prediction is defined as shown in Equation (1).

$$I = -\Sigma_i t_i (a_i - p_i) \log p_i \qquad (1)$$

where, for a given class *i*, $t_i$ is the probability that the neural network places the data in that class, $p_i$ is the fraction of data within the class, and $a_i$ is the accuracy of the prediction. Thus, Equation (1) calculates how much more information a prediction gives than a random guess based on the population of each class. The aim is to find an optimal combination of the information obtained from each prediction and the accuracy of this prediction. The information provided by each prediction when there are only two classes is shown in Figure 6.

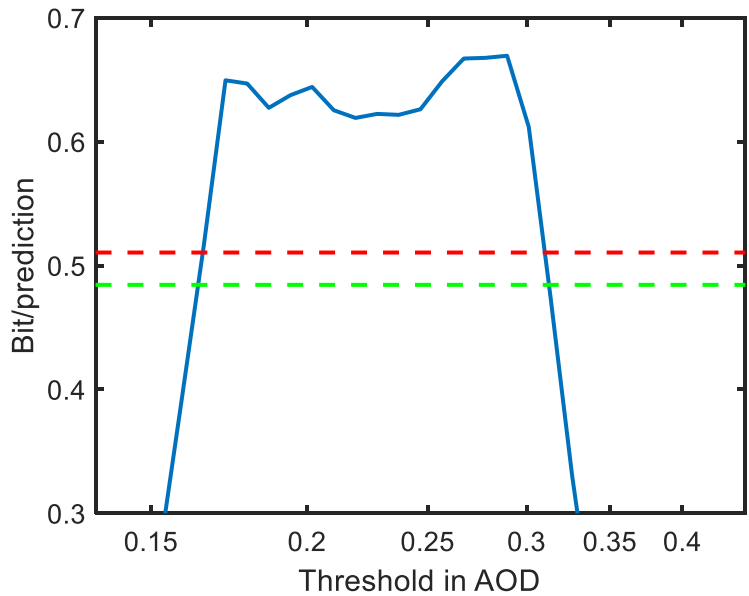

Figure 6. Relation between AOD threshold and information content for each prediction (the blue curve). The maximum information per prediction in the best 3-class and 4-class models are marked in the figure with red and green dashed lines, respectively.

It can be seen that AOD=0.3 serves as an optimal threshold delimiting the two classes; further, a two-class model outperforms models with more classes. The optimal structure of CF-NN is shown in Figure 7.

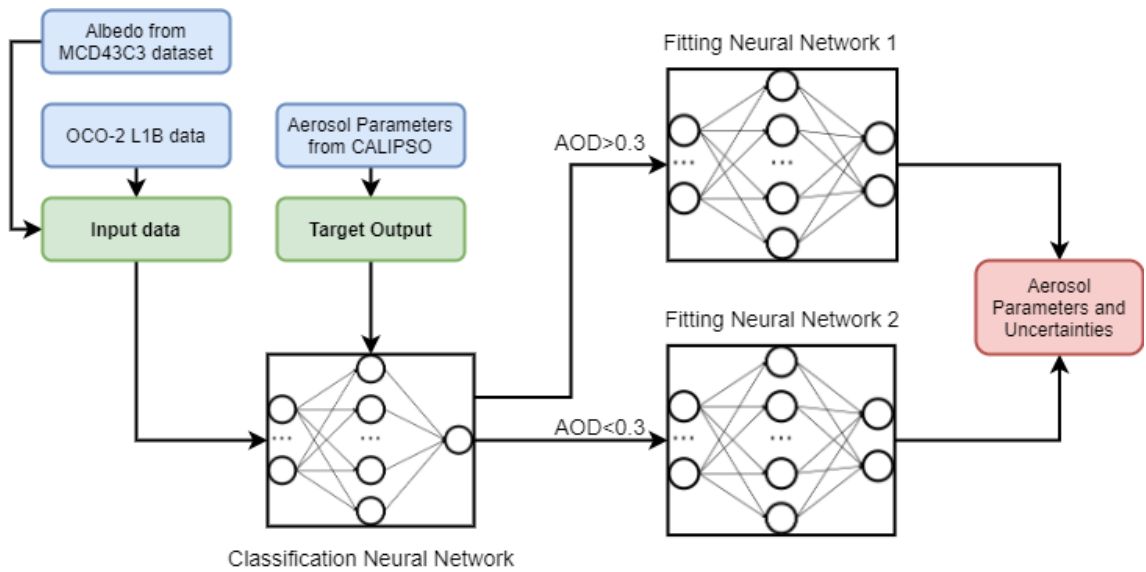

Figure 7. Structure of classification-fitting neural network (CF-NN). The classification neural network contains a hidden layer of 10 neurons and a single output as the result of a binary classification. The fitting neural network contains a hidden layer of 10 neurons and an output layer containing AOD or ALH with uncertainty estimations.

We use the maximum likelihood loss function for training, which allows us to evaluate the predicted values and their uncertainties at the same time. The same network structure and loss functions are used for both AOD and ALH fitting. In order to be consistent with the OCO-2 retrieval algorithm [11], AOD is predicted on a log scale, while ALH is predicted on a pressure scale. The maximum likelihood loss function uses the negative log-likelihood in its functional form, as shown in Equation (2). The ensembles are weighted in terms of the estimated error, as shown in Equation (3).

$$\mathcal{L} = \frac{\log \sigma^2(x)}{2} + \frac{(y - \mu(x))^2}{2\sigma^2(x)} \qquad (2)$$

$$\mu_*(x) = \frac{1}{N}\Sigma_i \mu_i(x), \sigma_*^2(x) = \frac{1}{N}\Sigma_i \left(\sigma_i^2(x) + \mu_i^2(x)\right) - \mu_*^2(x) \qquad (3)$$

The predicted AOD and ALH from the test dataset are used as *a priori* for the $CO_2$ retrieval. The retrieval with improved *a priori* is then compared with the OCO-2 Level-2 Standard data product and OCO-2 Level-2 Lite data product.

### 3.2 Error quantification

Given the absence of ground truth for $XCO_2$ in the target region, we conduct a simulation experiment to estimate the error in our prediction of AOD and ALH and subsequently in the retrieved $XCO_2$. We generate 4000 data points through forward radiative transfer simulations. The distribution of $XCO_2$ follows that of the Lite data, while those of albedo and AOD/ALH follow MCD43D3 and CALIPSO Level-2 data, respectively. Simulated observations of albedo and AOD/ALH are obtained by adding a gaussian noise with uncertainty being the mean estimated uncertainty in CALIPSO and MCD43D3 datasets. The relevant statistics from the simulation experiment are presented in Table 1.

Table 1 Statistics of key quantities from the simulation experiment

| Quantity | Data Source | Mean | Standard Deviation | Observation Error |
|---|---|---|---|---|
| AOD | CALIPSO Level-2 | 0.24 | 0.21 | 0.024 |
| ALH (km) | CALIPSO Level-2 | 2.07 | 0.85 | 0.72 |
| Albedo | MCD43D3 | 0.34 | 0.13 | 0.021 |
| $XCO_2$ (ppm) | OCO-2 Lite | 408.4 | 1.66 | None |

The forward simulations are conducted using the forward model in version 8 ACOS algorithm [8]. The data obtained are used for training and testing with the same machine learning methods. The outcome is compared with the error we obtained in comparison to the Lite data; we use this to estimate the quality of the Lite product.

## 4. Results
### 4.1 AOD and ALH retrieval

We first illustrate the performance of the AOD prediction. Results from the OCO-2 Level-2 Standard retrieval, the FC-NN, and the CF-NN algorithms are shown in Figure 8.

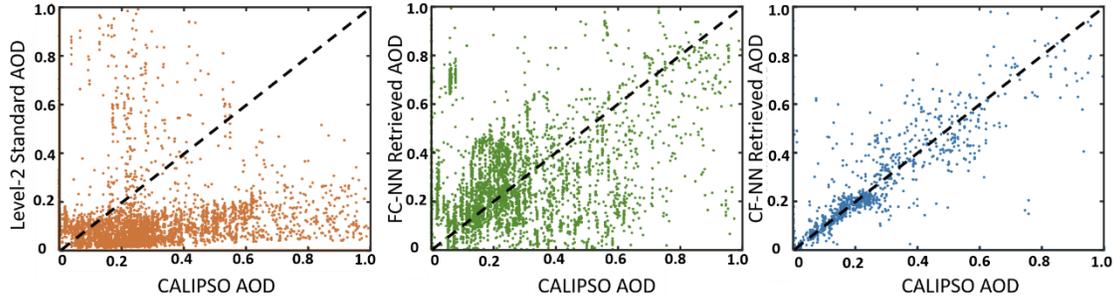

Figure 8. Comparison between retrieved $O_2$-A Band AOD from (left) Level-2 Standard, (middle) FC-NN, and (right) CF-NN algorithms and CALIPSO AOD data.

It is apparent that the OCO-2 Level-2 data product significantly underestimates the AOD for most of the scenarios. While FC-NN partially resolves this issue, the error is relatively large. Superior results are obtained with CF-NN. We evaluate the performance of the different algorithms using two metrics: mean square error (MSE) and Earth mover's distance (EMD). These two metrics are defined in Equations (4) and (5).

$$MSE = \frac{1}{M}\Sigma_i\big(y_i - \mu_i(\boldsymbol{x}_i)\big)^2 \qquad (4)$$

$$EMD = \int_{x_{min}}^{x_{max}} \left|\int_{x_{min}}^{x} f_1(x')dx - \int_{x_{min}}^{x} f_2(x')dx\right| dx \qquad (5)$$

Here, MSE is defined on a set of data and the corresponding predictions $y_i$ and $\mu_i$, while EMD is defined on a pair of distributions $f_1(x)$ and $f_2(x)$ and quantifies the difference between them. With CALIPSO AOD as the reference, the metrics for the three retrievals are shown in Table 2.

Table 2 Performance metrics for the three different retrieval methods

| Retrieval Method | Earth Mover's Distance | Mean Square Error |
|---|---|---|
| OCO-2 L2 Standard | 0.1501 | 0.0595 |
| FC-NN | 0.0631 | 0.0431 |
| CF-NN | 0.0031 | 0.0244 |

Clearly, CF-NN performs significantly better than the OCO-2 L2 and FC-NN retrievals. In particular, the EMD is significantly smaller, which means that the shape of the AOD distribution is captured by CF-NN.

While the Level-2 Standard data product contains the retrieved AOD, ALH is not provided; indeed, accurately retrieving ALH from passive remote sensing instruments such as CALIPSO and OCO-2 has been a difficult proposition. However, [10] has proved that spectral sorting is able to accurately quantify the ALH over ocean surfaces. For the area around Saudi Arabia analyzed in this work, the MSE of CF-NN results for ALH is 0.40 km$^2$, while the EMD is 0.075 km from CALIPSO ALH data. This error is much smaller than the uncertainty of the distribution (0.85 km), showing that the shape of the ALH distribution is captured very well by CF-NN.

The quality of uncertainty estimation from CF-NN is also evaluated using CALIPSO AOD and ALH data. For each sounding, we use the predicted uncertainty $\sigma_i$ to obtain a normalized error value $\tilde{\epsilon}_i$, defined in Equation (6). Ideally, the normalized error is expected to follow the normal distribution $N(0,1)$.

$$\tilde{\epsilon}_i = \frac{\mu_i - y_i}{\sigma_i} \qquad (6)$$

Figure 9 presents the distribution of normalized error for AOD and ALH. The fitted normal distributions for AOD and ALH are $N(-0.39, 1.17)$ and $N(-0.07, 0.86)$, respectively. The bias towards smaller values for AOD predictions can be attributed to the conservativeness of machine learning, which tends to underestimate the rare large AOD values. The closeness of the distributions of normalized error to a true normal distribution indicates that the aerosol statistics supplied as *a priori* for the XCO$_2$ retrieval are fairly accurate.

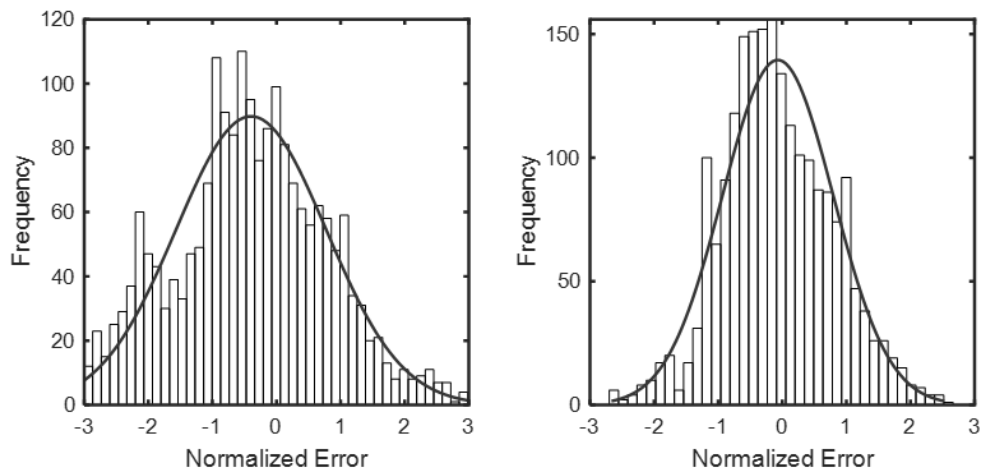

Figure 9. Distribution of normalized error in AOD (left) and ALH (right); the red line shows the fitted normal distribution.

## 4.2 XCO$_2$ retrieval

We use the AOD and ALH obtained above as the *a priori* constraint for the retrieval algorithm and compare with results from the Level-2 Standard algorithm. We also obtain results using AOD and ALH from CALIPSO as *a priori*. Previous studies have shown that CALIPSO-derived aerosol priors improve XCO$_2$ retrievals from OCO-2 [16]. Thus, we have two XCO$_2$ references, one from bias-corrected Level-2 Lite data, and the other from retrievals using CALIPSO aerosol data as *a priori*. In Figure 10, we show comparisons between the Level-2 Standard and machine learning-based retrievals and the two references (Level-2 Lite and CALIPSO-based). Table 3 presents the matrix of statistics, including standard deviation (Std) and bias, from these comparisons.

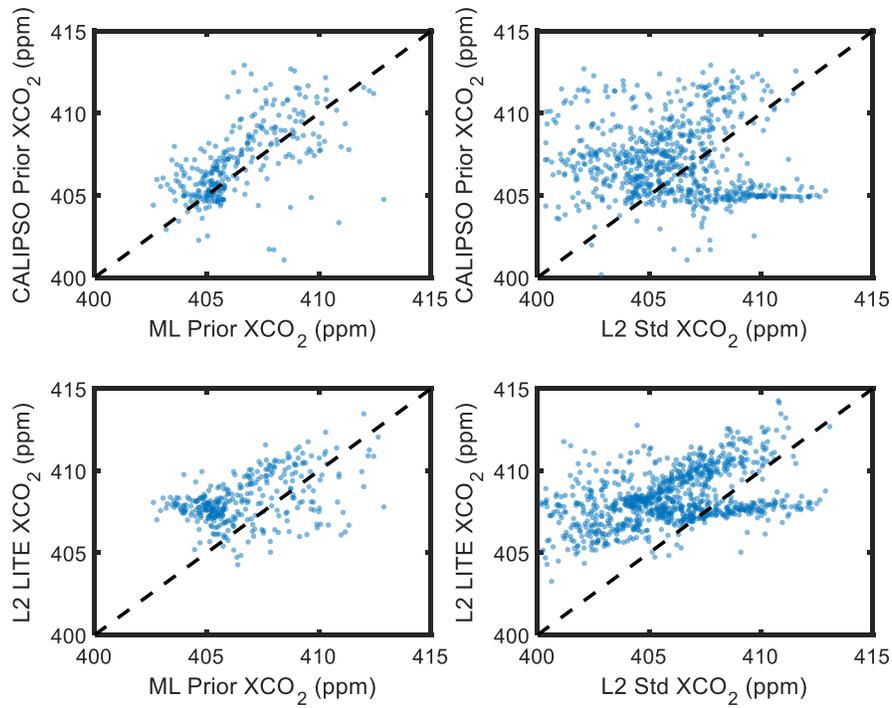

Figure 10. Level-2 Standard XCO$_2$ retrievals and retrievals with machine learning aerosol parameters as *a priori* compared to two references, Level-2 Lite XCO$_2$ and retrievals with CALIPSO aerosol parameters as *a priori*.

Table 3 Statistical comparison of $XCO_2$ retrieval results

|  | OCO-2 L2 Lite | | CALIPSO Aerosol Prior | |
|---|---|---|---|---|
| **OCO-2 L2 Standard** | Std (ppm) | Bias (ppm) | Std (ppm) | Bias (ppm) |
|  | 5.351 | -4.310 | 4.420 | -2.483 |
| **Machine Learning Prior** | Std (ppm) | Bias (ppm) | Std (ppm) | Bias (ppm) |
|  | 2.465 | -2.363 | 0.848 | -0.093 |

One important observation from Figure 10 is that the retrieval with machine learning prior agrees better with both sets of reference values. As shown in Table 3, both the standard deviation and bias are very small for the machine learning-based results, which is due to the high accuracy of predictions and a statistically accurate characterization of the uncertainties. In particular, the $XCO_2$ values retrieved with the machine learning prior are more compactly distributed. This agrees with our expectation, as significant $CO_2$ sources or sinks are absent over the desert region, and the concentrations should be homogeneous.

However, when OCO-2 L2 Lite data are used as the reference, although the machine learning results are an improvement over the L2 Standard retrieval, the error and bias are larger than when we use CALIPSO-based priors as the reference. It then becomes necessary to determine which reference is more robust. While CALIPSO aerosol information potentially contains errors, the bias correction for Lite data is global and is possibly biased on regional scales. To evaluate the accuracy of the machine learning methodology, we employ the simulations described in the previous section.

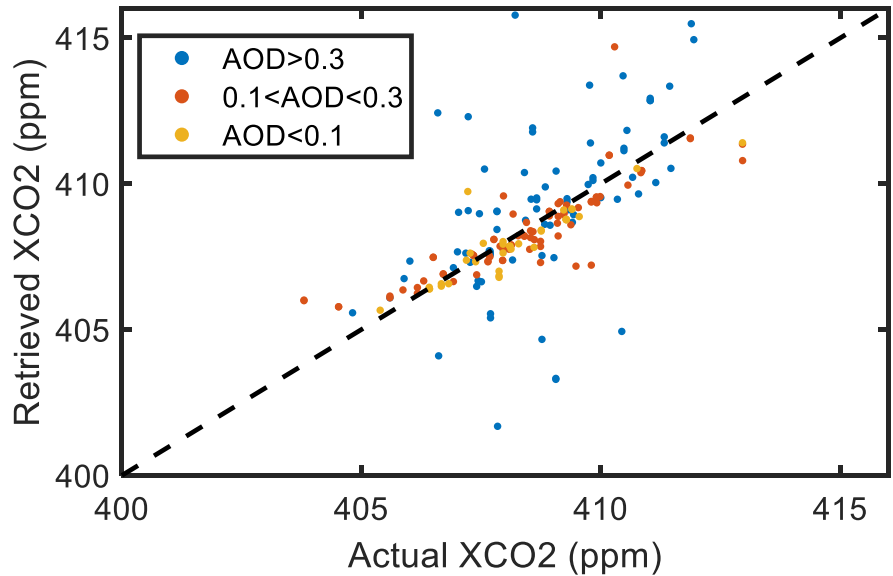

Figure 11. Relation between retrieved and true XCO$_2$ used in the simulation experiment

The results of the simulation experiment are shown in Figure 11. The data points are color-coded by their respective AOD values. The aerosol loading has a significant influence on the accuracy of prediction. When AOD is above 0.3, the scatter of the data points is significantly larger than that for values below 0.3.

The overall standard deviation of the retrieved XCO$_2$ is 1.79 ppm, and the bias is -0.04 ppm, which is negligible. The algorithm is robust to albedo observation errors: increasing the albedo observation error to 0.10 and 0.15 increases the XCO$_2$ Std to 1.93 and 2.29 ppm, respectively. On the other hand, if the AOD error is increased to 0.05, the XCO$_2$ Std increases to 3.50 ppm. However, for all the cases, no significant bias is observed. Hence, we conclude that we are more confident in the retrieval with machine learning prior, and the OCO-2 Lite data may be biased around the Saudi Arabia region.

## 5.Conclusion

In this work, we retrieve aerosol information from OCO-2 spectra with a classification-fitting neural network method. Our results demonstrate that the two-step machine learning model provides greatly improved AOD and ALH results compared with the classical FC-NN model. A spectral-sorting-based data reduction method facilitates application of the machine learning technique on large volumes of observational data. Both the EMD and MSE statistical metrics are significantly reduced, compared with L2 Standard data, for AOD and ALH, indicating that an improved fit is achieved.

Uncertainty prediction is performed using a maximum likelihood loss function. The predictions are used as *a priori* for $XCO_2$ retrievals; the machine learning retrievals are shown to be more accurate compared with L2 Standard data. By conducting a simulation experiment, the error of the machine learning method is quantified, including for albedos near the critical value. We have further shown that the large bias in comparison to L2 Lite data is unlikely to be due to errors in the machine learning methodology. Rather, the globally bias-corrected Lite $XCO_2$ values may be inaccurate in some regions.

This method has potential application for a wide range of surface types. On the other hand, effects of clouds have not been considered; accuracy of predictions could be reduced by cloud coverage. We expect that future improvements will allow the application of these methods to regions with thin clouds such as high-altitude cirrus.

## Acknowledgement

A portion of this research was carried out at the Jet Propulsion Laboratory, California Institute of Technology, under a contract with the National Aeronautics and Space Administration (80NM0018D0004). VN acknowledges support from the NASA Earth Science US Participating Investigator program (solicitation NNH16ZDA001N-ESUSPI). We thank D. Crisp, C. E. Miller, M. Gunson and S. Sander for stimulating discussions.

## Author Contributions

S. C. devised the machine learning method, conducted the data curation, performed the analysis, and wrote the manuscript. V. N. conceived the project, supervised the work, and revised the manuscript. Z. Z. developed the spectral sorting method, revised the manuscript, and provided suggestions on visualization. Y. Y. co-supervised the work and revised the manuscript.